# Probing the microscopic structure and flexibility of oxidized DNA by molecular simulations

K B Chhetri[1,2], S Naskar[1] and P K Maiti[1]*

[1]Department of Physics, Center for Condensed Matter Theory, Indian Institute of Science, Bangalore 560012, India

[2]Department of Physics, Prithvinarayan Campus, Tribhuvan University, Pokhara, Nepal

**Abstract:** The oxidative damage of DNA is a compelling issue in molecular biophysics as it plays a vital role in the epigenetic control of gene expression and is believed to be associated with mutagenesis, carcinogenesis and aging. To understand the microscopic structural changes in physical properties of DNA and the resulting influence on its function due to oxidative damage of its nucleotide bases, we have conducted all-atom molecular dynamic simulations of double-stranded DNA (dsDNA) with its guanine bases being oxidized. The guanine bases are more prone to oxidative damage due to the lowest value of redox potential among all nucleobases. We have analyzed the local as well as global mechanical properties of native and oxidized dsDNA and explained those results by microscopic structural parameters and thermodynamic calculations. Our results show that the oxidative damage of dsDNA does not deform the Watson-Crick geometry; instead, the oxidized DNA structures are found to be better stabilized through electrostatic interactions. Moreover, oxidative damage changes the mechanical, helical and groove parameters of dsDNA. The persistence length, stretch modulus and torsional stiffness are found to be 48.87 nm, 1239.26 pN and 477.30 pN.nm$^2$, respectively, for native dsDNA and these values are 61.31 nm, 659.91 pN and 407.79 pN.nm$^2$, respectively, when all the guanine bases of the dsDNA are oxidized. Compared to the global mechanical properties, the changes in helical and groove properties are found to be more prominent, concentrated locally at the oxidation sites and causing the transition of the backbone conformations from BI to BII at the regions of oxidative damage.

## 1. Introduction

DNA oxidation, the oxidative damage of deoxyribonucleic acid, is being the subject of interest due to the biological repercussions like genome instability and mutation that it brings to our body [1]. Among four DNA nucleobases, guanine is more prone to oxidation since it has the lowest redox potential [2, 3]. During oxidative phosphorylation, reactive oxygen species (ROS) are generated in the form of super-oxides ($O_2^-$) and $H_2O_2$, which are responsible for oxidation of biomolecules like protein, DNA, etc. [4]. ROS can also be produced by ionizing or UV radiation. ROS may interact with biological macromolecules such as DNA, causing alteration and possibly severe repercussions to the cell, regardless of their source (endogenous or exogenous) [5]. The DNA alterations in mammalian chromatin caused by free radicals are found to be associated with mutagenesis, carcinogenesis and aging [6–9].

Modification of DNA bases is not only malignant but also plays a vital role in the epigenetic control of gene expression [10]. Practically, each of the DNA bases can be modified; however, modifications of guanine and adenine bases are most occurring due to their smaller redox potentials [11]. The 7,8-dihydro-8-oxoguanine, also known as 8-oxoguanine (8oxoG), is one of the most abundant byproducts of oxidative DNA damage [12]. In addition to being an output of DNA oxidative damage, 8oxoG has a role in transcriptional regulation under oxidative stress [13]. During hypoxia, intra cellular degrees of ROS are raised, which facilitates the production of 8oxoG [14, 15]. Excess oxidative DNA damage is linked to cancers and many other diseases, while a small amount of oxidized nucleotides produced due to normal ROS levels is necessary for memory and learning [16–18].

*Corresponding author, E-mail: maiti@iisc.ac.in

In 2016, while studying the effect of cytosine modifications on DNA flexibility, Ngo et al. [19] found that the methylation of cytosine (5-methylcytosine) causes reduction of DNA flexibility. During base analog substitution (substituting either inosine for guanosine or 2,6-diaminopurine for adenine), Peters et al. [20] found minor changes in global properties like persistence length, helical repeat and torsional stiffness. The data obtained from circular dichroism spectroscopy in the same work of Peters et al. showed some significant changes in helical geometry of the modified DNA compared to normal DNA. In another work, Peters et al. [21] experimentally measured the bending and twisting flexibilities of DNA analog polymers with one of the four regular bases of DNA substituted by different cationic, anionic, or neutral analogs under low salt buffers. They found only about 20% change in bending rigidity but a large increase (about 5-fold) of twist flexibility on such modified DNA analogs in comparison to the unmodified one. They suggested that such modifications of regular bases make dsDNA to have transition to different helical conformations other than canonical B-form and effect is minimal as far as the mechanical properties are concerned. It is reported that the methylation of the cytosine base of DNA influence the DNA's backbone structure due to the steric hindrance between the methyl group and ribose sugar that prevents the formation of hydrogen bonds between the nucleobase and backbone and results in the local increment of DNA flexibility [22].

In another computational study by Miller et al. [23], G19:C6 base pair of DNA oligonucleotide GGGAA-**C**AACTAG:CTAGTT**G**TTCCC was replaced by 8oxoG. They found that when 8oxoG replaced G19, the local bending into the major groove is more probable than changing the DNA's global bending, which assists in the formation of local kinks at the 8oxoG associated major grooves. The oxidative damage of DNA can bring local alterations on phosphate backbone and changes of sugar puckers of the oxidized bases [24, 25]. Cheng et al. [26] conducted unrestrained molecular dynamics simulations for several 13-mer DNA duplexes. In their results, the B-form duplexes of oligomers with G:C and 8oxoG:C base pairs are found to adopt proper Watson-Crick geometry and the local and global flexibilities of the duplexes are increased. In the case of G:A mismatch, the Watson-Crick geometry is found to be decreased with higher structural fluctuations. These simulations demonstrated that both dynamic and equilibrium properties of DNA duplexes change during their oxidative damage. All these works inspired us to decipher how different amount of oxidation in DNA can influence their microscopic structural and mechanical properties.

In this work, we carried out all-atom MD simulations of different oxidized double-stranded DNAs (dsDNAs) and computed various mechanical properties such as stretch modulus ($\gamma_G$), persistence length ($l_p$), twist-stretch-coupling ($\tau$) and torsional stiffness (C). As several earlier works have shown more alterations of local parameters than global ones, we have tried to explain the changes of various microscopic structural parameters with the oxidization of the DNA bases. We believe, it will help to advance the understanding about the alterations of biological as well as physical properties brought due to oxidative damage of DNA, specifically based on the changes of mechanical properties and microscopic helicoidal parameters.

This article is organized as follows. We begin with the methods describing the model building of oxidized DNA and details of all-atom MD simulations. Then, we give details of the theoretical models used to calculate the elastic properties of nucleic acids. In the results and discussion section, we present the mechanical properties, such as stretch modulus ($\gamma_G$), persistence length ($l_p$), twist-stretch-coupling ($\tau$), etc., of the dsDNAs. We also analyze the various microscopic structural parameters. Finally, we summarize all the results and provide a perspective and utility of various results.

## 2. Materials and methods

### 2.1. Simulation setup

#### 2.1.1. System build-up

The Dickerson-Drew dodecamer (d[CGCGAATTCGCG]) double-stranded DNA (dsDNA) was prepared using the nucleic acid builder(NAB) [27] tool of Amber18 [28]. Using an in-house developed python script, the guanine bases are oxidized as shown in Fig. 1(b). We prepared three dsDNA molecules: one is native dsDNA, where no bases are oxidized, another one is dsDNA(4oxG), whose four guanine bases (two of each strand but in two opposite halves) are oxidized and the third one is dsDNA(8oxG), whose eight guanine bases (four of each strand or all guanine bases) are oxidized.

The interactions of pure dsDNA are represented by Amber ff10 force field [29] while the interactions of oxidized bases are taken from the work of Miller et al. [23]. The TIP3P water model [30, 31] was used to solvate the dsDNAs, resulting in a 15 Å TIP3P water buffer surrounding the structure in each direction. Because the

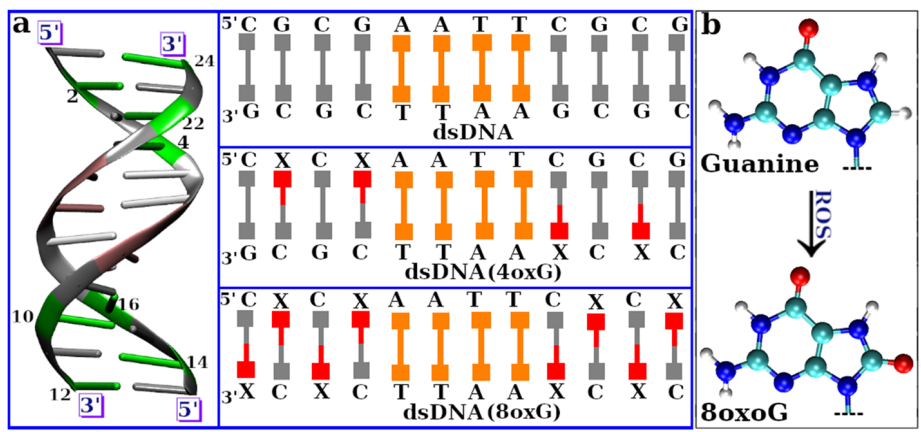

**Fig. 1** (a) Schematics for the dsDNA duplexes with X as an oxidized guanine. The dsDNA represents the DNA duplex with none of the bases are oxidized, dsDNA(4oxG) represents the DNA duplex whose four guanine bases (two of each strand but on opposite halves) are oxidized and dsDNA(8oxG) represents the DNA duplex whose eight guanine bases (four of each strand or all guanines) are oxidized. (b) Conversion of guanine to oxoguanine (8oxoG) with its oxidative damage. The larger dark silver spheres, smaller bright silver spheres, blue spheres and red spheres represent carbon, hydrogen, nitrogen and oxygen atoms, respectively phosphate backbones in dsDNA are negatively charged, the systems needed to be neutralized, which was accomplished by introducing the necessary amount of $Na^+$ ions. The Joung-Cheatham ion parameter set was used to characterize the interaction of ions with water and nucleic acids [32]. In this way, three different solvated systems are prepared, corresponding to dsDNA, dsDNA(4oxG) and dsDNA(8oxG), respectively. Schematically these three dsDNA duplexes are described by Fig. 1(a).

### 2.1.2. MD simulation methodology

To remove any bad contacts that arise during the system preparation, we performed energy minimization of systems. We used the steepest descent algorithm (for 2500 steps), followed by the conjugate gradient algorithm (next 2500 steps). All of the solute atoms (nucleic acid atoms) were kept fixed, applying a harmonic potential of spring constant 500 kcal.mol$^{-1}$Å$^{-2}$. The restraint applied to the solute atoms was reduced to zero in five stages with 5000 steps of energy minimization in each stage. That means the positional restraint applied to solute atoms was made zero during the last 5000 steps of equilibration. The energy minimized systems were then heated from 10 K to 300 K in four steps: 10K to 50K, 50 K to 100 K, 100 K to 200 K and 200 K to 300 K. The dsDNA was position restrained using a harmonic constant of 20 kcal.mol$^{-1}$Å$^{-2}$ during the whole heating process. We used Langevin thermostat [33, 34] with a coupling constant of 0.5 ps to control the temperature. To equilibrate, the systems were subjected to a 2 ns NPT simulation following the heating. Berendsen weak coupling method [35, 36] with a coupling constant of 0.5 ps was employed to maintain the pressure to 1 atm. At last, 500 ns long MD simulation was conducted in the NVT ensemble with 2 fs integration time steps. During simulation, the SHAKE algorithm [37] was adopted to constrain the hydrogen bonds. To account for the electrostatic interactions, we used the Particle Mesh Ewald (PME) method [38]. We used an LJ potential cut-off of 10 Å. At the cut-off, the van der Waals (vdW) and direct electrostatic interactions were terminated. Similar simulation methodologies have been successfully implemented in several of our previous studies involving DNA and DNA-based nanostructures [39–44].

## 2.2. Theories on mechanical properties

### 2.2.1. Persistence length

In DNA, the persistence length is the elementary section of its length over which the correlations in the tangent directions are lost. The persistence length of DNA measures its bending rigidity. Molecules having larger persistence lengths are considered to be stiffer to bend. There exist many theoretical models to compute the persistence length ($l_p$) of dsDNA, described in the article of Garai et al. [41]. One of the ways of calculating ($l_p$) is through bending angle distribution. If $\vec{t}_1$ be the unit vector to the first base pair and $\vec{t}_n$ be the unit vector to the $n^{th}$ base pair then the bending angle ($\theta$) of dsDNA is defined as, $\theta = cos^{-1}(\vec{t}_1 \cdot \vec{t}_n)$.

Figure 2 describes end-to-end distance ($L_e$), contour length ($L$) and bending angle ($\theta$) of dsDNA shown schematically. If $l_i$ be the center to center distance between two consecutive base pairs then the average contour length is defined as: $L_0 = <\sum_{i=1}^{n} l_i>$, where $< >$ is the time average over all frames.

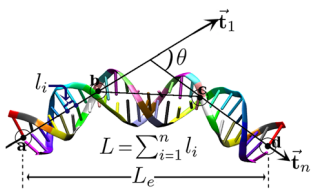

**Fig. 2** Schematic representation for end-to-end distance ($L_e$), contour length (L) and bending angle ($\theta$) of dsDNA. The bending angle ($\theta$) is the angle between the unit vectors $\vec{t}_1$ and $\vec{t}_n$ of the terminal base pairs

For dsDNA, the bending angle distribution can be approximated by Gaussian distribution [45–47] as follows:

$$P(\theta) = \sqrt{\frac{\beta\kappa}{2\pi L_0}} e^{-\frac{\beta\kappa}{2L_0}\theta^2} \quad (1)$$

For small angle $\theta$, equation 1 can be approximated as [45, 48]:

$$lnP(\theta) = -\frac{l_p}{L_0}(1-cos\theta) + \frac{1}{2}ln\left(\frac{\beta\kappa}{2\pi L_0}\right) \quad (2)$$

where $\kappa = \frac{l_p}{\beta} = K_B T l_p$ is the bending modulus, $T$ is temperature of the system in Kelvin and $K_B$ is the Boltzmann's constant. From the slope of the graph of $lnP(\theta)$ versus $(1-cos\theta)$, we can compute the persistence length using: $slope = -\frac{l_p}{L_0}$.

### 2.2.2. Stretch modulus

The stretch modulus of DNA gives the measure of its stretching rigidity. Molecules having greater stretch modulus are considered to be stiffer to stretch. To calculate the stretch modulus, we computed the contour length distribution of the dsDNA. The contour length distribution also follows the Gaussian distribution, given by [41, 49]:

$$P(L) = \sqrt{\frac{\beta\gamma_G}{2\pi L_0}} e^{-\frac{\beta\gamma_G}{2L_0}(L-L_0)^2} \quad (3)$$

Equation 3 can be re-expressed as [49, 50]:

$$lnP(L) = -\frac{\beta\gamma_G L_0}{2}\left(\frac{L}{L_0} - 1\right)^2 + \frac{1}{2}ln\frac{\beta\gamma_G}{2\pi L_0} \quad (4)$$

where $\gamma_G$ is the stretch modulus. From the slope of the graph of $lnP(L)$ versus $\left(\frac{L}{L_0} - 1\right)^2$, we can compute the stretch modulus ($\gamma_G$).

### 2.2.3. Twist-stretch coupling

Twist-stretch coupling ($\tau$) is the correlation between the helical-rise (H-rise) and helical-twist (H-twist) of the double-stranded nucleic acids. It shows the correspondence between the twisting and stretching rigidity of the molecule under its deformation [51]. A positive value of $\tau$ implies that the molecule stretches with its twisting and a negative value of $\tau$ means that the twisting of the molecule causes lateral dilation, increasing the inter-strand distance and decreasing the length [52].

The twist-stretch coupling ($\tau$) is associated with the elements of covariance matrix that are related with H-rise and H-twist as: [53, 54],

$$\tau = \frac{cov(L, \Phi)}{cov(L, L)} = \frac{d(H\text{-}rise)}{d(H\text{-}twist)} \quad (5)$$

where $cov(L, L)$ and $cov(L, \Phi)$ are the variance of $L$ and covariance between $L$ and $\Phi$, respectively, such that $L$ is the sum of H-rise parameter and $\Phi$ is the sum of H-twist parameter of each base pair step.

Hence, the twist stretch coupling is given by: d(H-rise)/d(H-twist), the slope of the straight line obtained through the regression fitting of H-rise and H-twist parameters.

### 2.2.4. Torsional Stiffness

The mechanical property of dsDNA which is associated with the twisting stiffness against torsional deformation is called its torsional stiffness or twisting rigidity (C). The torsional stiffness of dsDNA is also associated with the diagonal element of covariance matrix corresponding to H-twist parameter or twist angle ($\Phi$) and is given by: [55, 56]

$$C = \frac{K_B T L}{\sigma_\Phi^2} \quad (6)$$

where $\sigma_\Phi^2$ is the variance of angle $\Phi$. As before, $L$ is the sum of H-rise parameter and $\Phi$ is the sum of H-twist parameter of each base pair step. The twist related persistence length is given by: $\frac{L}{\sigma_\Phi^2}$ [51, 57].

### 2.2.5. Crookedness

Crookedness of dsDNA quantifies its curvature and signifies how easily it can extend and deviate from its helical axis. It is characterized by a parameter ($\beta$) which is defined as [58]:

$$cos\beta = \frac{L_e}{\sum l_i} \quad (7)$$

where $\sum l_i$ is the contour length, the sum of center to center distances of consecutive base pairs and $L_e$ is the end-to-end distance. A DNA duplex can elongate by reducing its curvature $\beta$. The duplex has a greater stretch modulus only if its $\beta$ value is small [58].

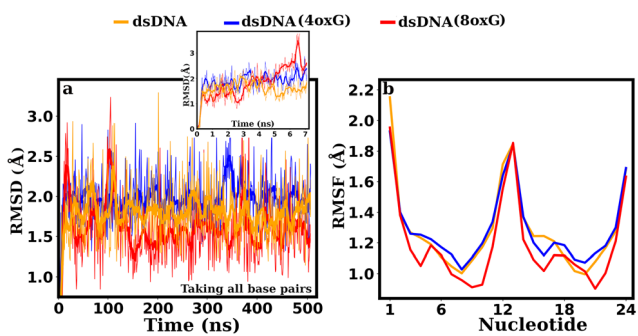

**Fig. 3** (**a**) RMSD plots and (**b**) RMSF plots for the three Dickerson-Drew dsDNA molecules. Where dsDNA represents native dsDNA with none of the bases oxidized, dsDNA(4oxG) represents dsDNA with four guanines (two of each strand but in opposite halves) oxidized and dsDNA(8oxG) represents dsDNA with eight guanines (four in each strand) oxidized. For distinct appearance, an inset plot for RMSD is also included which shows initial variations of RMSD more clearly

The trajectories of the simulated duplexes were analyzed using the 3DNA software [59], VMD [60] and CPPTRAJ [61] tool provided with Amber2018 software [28].

## 3. Results and discussion

In order to confirm that the simulated dsDNA structures are stable during the MD simulation, we computed root mean square deviation (RMSD) and root mean square fluctuation (RMSF) of the dsDNA molecules with respect to their equilibrated structures.

The root mean square deviation (RMSD) plots in Fig. 3(a) show that both non-oxidized dsDNA and oxidized dsDNAs are stable during the course of the simulations.

For dsDNA and dsDNA(4oxG), RMSD values are not much different except at around 350 ns. For dsDNA(8oxG), RMSD is relatively lower than others. The relatively lower RMSD values for dsDNA(8oxG) imply that the dsDNA is stabilized better with the increase in the number of oxidation sites in it. From the root mean square fluctuation (RMSF) plots shown in Fig. 3(b), we observe that, rather than globally, there are some structural changes on the oxidized dsDNA locally at the sites of oxidative damage. We also observe that terminal bases fluctuate more and have higher RMSF compared to the central bases.

### 3.1. Mechanical properties

The calculated values of various mechanical properties of all the three dsDNA molecules, namely; persistence length, stretch modulus, torsional stiffness and twist-stretch coupling, are given in Table 1.

The persistence length ($l_p$) of the dsDNAs were computed using equation 2. The linear fit plots to equation 2 are included in Fig. SF.1[B] of supplementary information (SI). When the dsDNA is not oxidized, the value of $l_p$ is $48.87 \pm 1.81$ nm which is very close to the experimentally measured persistence lengths of ~45-50 nm for dsDNA in short length scales [53, 62–65]. With oxidation of bases, the $l_p$ is increased to $58.61 \pm 2.06$ nm and $61.31 \pm 2.44$ nm for dsDNA with four guanines oxidized and for dsDNA with all of the eight guanines oxidized, respectively, signifying that the increase of oxidation sites increases the $l_p$ of dsDNA that restrains the bending flexibility. In the SI, we have computed the $l_p$ for the duplexes taking their intrinsic curvature as well (see Table St.1 of SI) and with that, the values of $l_p$ are slightly changed from the values

**Table 1** Different mechanical properties for the systems of simulation

| System | Persistence Length ($l_p$) (nm) | Stretch Modulus ($\gamma_G$) (pN) | Torsional Stiffness(C) (pNnm$^2$) | Twist-Stretch Coupling($\tau$) (nm/rad) |
| --- | --- | --- | --- | --- |
| dsDNA | 48.87 ± 1.81 | 1239.26 ± 80.47 | 477.30 ± 11.16 | 0.157 ± 0.001 |
| dsDNA(4oxG) | 58.61 ± 2.06 | 800.68 ± 43.76 | 424.51 ± 7.21 | 0.221 ± 0.001 |
| dsDNA(8oxG) | 61.31 ± 2.44 | 659.91 ± 35.82 | 407.79 ± 13.32 | 0.103 ± 0.001 |

*dsDNA: No bases are oxidized, dsDNA(4oxG): four guanine bases are oxidized, dsDNA(8oxG): All of the eight guanine bases are oxidized

**Table 2** Contour length, end-to-end distance, bending angle and crookedness of various duplexes

| System | Contour Length($L_0$) (Å) | End-to-end Distance ($L_e$) (Å) | Bending Angle ($\theta$) (deg) | Crookedness ($\beta$) |
| --- | --- | --- | --- | --- |
| dsDNA | 36.46 ± 0.64 | 36.09 ± 0.93 | 20.67 ± 12.14 | 0.18 ± 0.07 |
| dsDNA(4oxG) | 36.38 ± 1.13 | 35.37 ± 1.27 | 20.30 ± 10.30 | 0.25 ± 0.08 |
| dsDNA(8oxG) | 36.22 ± 1.71 | 32.82 ± 1.37 | 19.42 ± 10.12 | 0.43 ± 0.11 |

*As an abbreviation, degree ($^o$) is written as deg throughout this article

**Table 3** Averages of some of the helicoidal parameters of various DNA duplexes studied in this work

| System | H-rise (Å) | Inclination (deg) | H-twist (deg) | Roll (deg) |
| --- | --- | --- | --- | --- |
| dsDNA | 3.32 ± 0.41 | 6.58 ± 12.31 | 35.55 ± 6.51 | 2.96 ± 8.64 |
| dsDNA(4oxG) | 3.16 ± 0.52 | 7.80 ± 15.06 | 34.08 ± 7.47 | 3.89 ± 7.38 |
| dsDNA(8oxG) | 2.74 ± 0.68 | 14.32 ± 19.68 | 31.20 ± 10.86 | 4.17 ± 6.86 |

*The averages are taken from the values of the respective parameters of guanine-involved base pairs. Few more helicoidal parameters are presented in Table 4

presented in Table 1 but follow the same trend. The decrease in average and standard deviation values of the bending angle presented in Table 2 shows that dsDNA becomes less prone to bending with the increase in the number of the oxidation sites, which substantiates that the oxidative damage of dsDNA makes it relatively stiffer to bend in comparison to the native dsDNA.

The stretch modulus ($\gamma_G$) of the dsDNAs were computed using equation 4. The linear fit plots to equation 4 are included in Fig. SF.1[A] of SI. Without oxidation of bases, the $\gamma_G$ is 1239.26 ± 80.47 pN which agrees with the experimental results ∼1000-1500 pN of dsDNA in short length scales [53, 63, 66, 67]. With oxidation of bases, the $\gamma_G$ is decreased to 800.68 ± 43.76 pN and 659.91 ± 35.82 pN, for dsDNA with four guanines oxidized and for dsDNA with all of the eight guanines oxidized, respectively, signifying that the increase of oxidation sites decreases the $\gamma_G$ of dsDNA thereby making it more flexible to stretch in comparison to the native dsDNA. The higher values of standard deviations of contour length and end-to-end distance, as well as the higher values of crookedness presented in Table 2, also show that dsDNA becomes more flexible to stretch with the increase in the number of oxidation sites. We found the smaller value of stretch modulus for the duplexes with larger crookedness. That means with the increase in crookedness molecule becomes less stiff to stretch.

The torsional stiffness or twist modulus (C), computed using equation 6, is 477.30 ± 11.16 pN.nm² without oxidation of the dsDNA. It decreased with increase in the number of oxidized bases and became 424.51 ± 7.21 pN.nm² and 407.79 ± 13.32 pN.nm² when four guanines are oxidized and when all guanines are oxidized, respectively. It signifies that as the number of oxidation sites increases, the dsDNA becomes less rigid in terms of stretching and twisting. The high flexibility of oxidized dsDNA to stretch and twist are related to its relatively open (laterally stretched) structure, described in the next section.

We have studied the twist-stretch coupling of the duplexes to know how their extension is related to their twisting. For twist-stretch coupling, its sign is important since it provides information about the geometry of the structure. A-form geometries of dsDNA or dsRNA have twist-stretch couplings opposite (in sign) to their B-form geometries [52, 67]. Positive twist-stretch coupling implies the direct correlation between helical twist (H-twist) and helical rise (H-rise) parameters of the duplex whereas, negative twist-stretch coupling implies the inverse correlation between them. Here, for all three dsDNAs, the twist-stretch coupling ($\tau$) is positive (see Table 1). It asserts the direct correlation between twist and stretch that with increase/decrease in helical-twist, the helical-rise should increase/decrease accordingly.

From the mechanical properties computation, we found that the oxidized B-form dsDNA shows its mechanical properties closure toward that of the A-form duplexes. For A-form duplex, the persistence length is greater than that of B-from duplex, whereas the stretch modulus and twist modulus of A-form duplex are less significant than those of its B-form duplex. For instance, the mechanical properties data of A-form duplex can be referenced from that of A-form dsRNA. The available experimental values of the persistence length are ∼45-50 nm and ∼60 ± 10 nm [53, 62–65] for B-form dsDNA and A-form dsRNA duplexes, respectively. The available experimental values of stretch modulus are ∼1000-1500 pN and ∼350–600 pN [53, 63, 66, 67] for B-form dsDNA and A-form dsRNA duplexes, respectively. The experimental values of torsional persistence length in the research work by Lipfert et al. [67] are 109 ± 4 nm for B-form dsDNA and 100 ± 2 nm for A-form dsRNA corresponding to torsional stiffness or twist modulus 451.48 ± 16.57 pN.nm² and 414.20 ± 8.28 pN.nm², respectively.

The mechanical properties computation results in this work (oxidized B-form dsDNA showing its mechanical properties closure toward that of the A-form duplex) agree with the work of Peters et al. [21] that the more dominant effect of nucleotide bases modifications is the transition of helical conformations somewhat different from canonical B-form. The modifications brought on dsDNA bases by their oxidation change [19] and this change is minor on bending, but more in a twist or torsion [21]. Similar trend can be seen in our results, presented in Table 1.

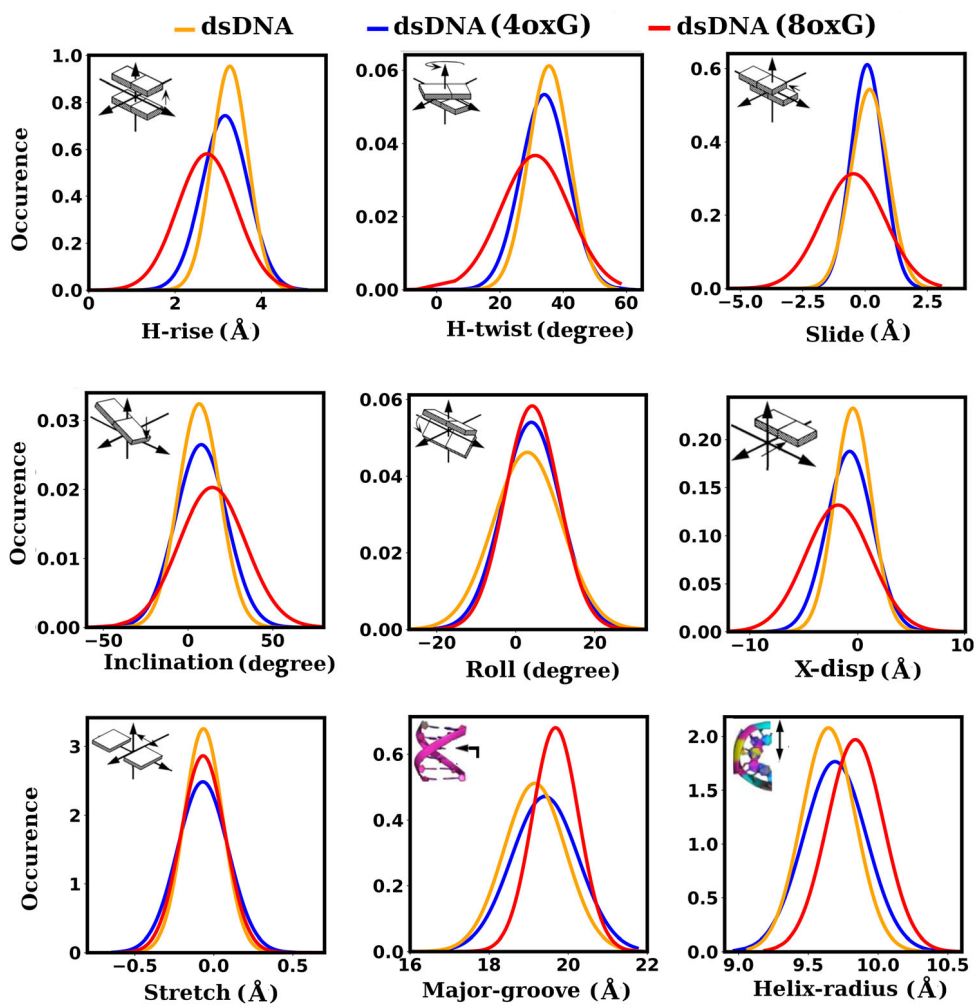

**Fig. 4** Average values of helicoidal and groove parameters taking CG base pairs only. The schematics of the corresponding parameters at the upper-left corners of the plots are taken from the ref. [68]

### 3.2. Helicoidal and groove parameters

To show how the helicoidal geometry of dsDNA changes with oxidation of its base pairs, we have computed some of the important helicoidal parameters of the dsDNA systems and are presented in Table 3.

The helical rise (H-rise) of dsDNA is the vertical distance between two consecutive base pairs. It is the projection of the rise (distance between the center of masses of two consecutive base pairs) parameter along the helical axis. We found the H-rise to be $3.32 \pm 0.41$ Å for the dsDNA without oxidation. With oxidation of the bases, the value of H-rise decreases. The decrement becomes more prominent with the increase in the number of oxidation sites. For the dsDNA, whose all bases are oxidized, the H-rise value is found to be $2.74 \pm 0.68$ Å. The helical-twist (H-twist) is the angle of rotation between neighboring nucleotide base pairs. The H-twist is found to be $35.55 \pm 6.51^o$ for the non-oxidized dsDNA and is decreased to $31.20 \pm 10.86^o$ when all guanine bases are oxidized. The inclination is the angle by which a base pair is inclined to the horizontal plane. The average value of inclination is obtained to be $6.58 \pm 12.31^o$ for the case in which no nucleotide bases of dsDNA are oxidized. Unlike H-rise and H-twist, the inclination is found to increase with an increase in the oxidation sites in dsDNA and became $14.32 \pm 19.68^o$ when all guanine bases are oxidized. The effect of an increase in the inclination ($\alpha$) parameter can be seen in the decrease of the H-rise parameter since they are related as; $H\text{-}rise = rise \times cos\alpha$.

In general, the A-forms of dsDNA and dsRNA have smaller values of H-twist and H-rise than the B-form dsDNA [69]. The angle of inclination is greater for the A-forms of dsDNA and dsRNA than the B-form dsDNA [70]. The values of H-rise, H-twist and inclination presented in Table 3 show the similar trend that the oxidized dsDNAs have lower values of H-rise and H-twist and greater value of inclination than that of non-oxidized dsDNA. However, we can say that none of the duplexes do have transitions to A-form as the twist-stretch coupling is positive (see Table 1) for all of them and the sugar puckers at all base pairs including the oxidized ones (8oxoG-C pairs) appear

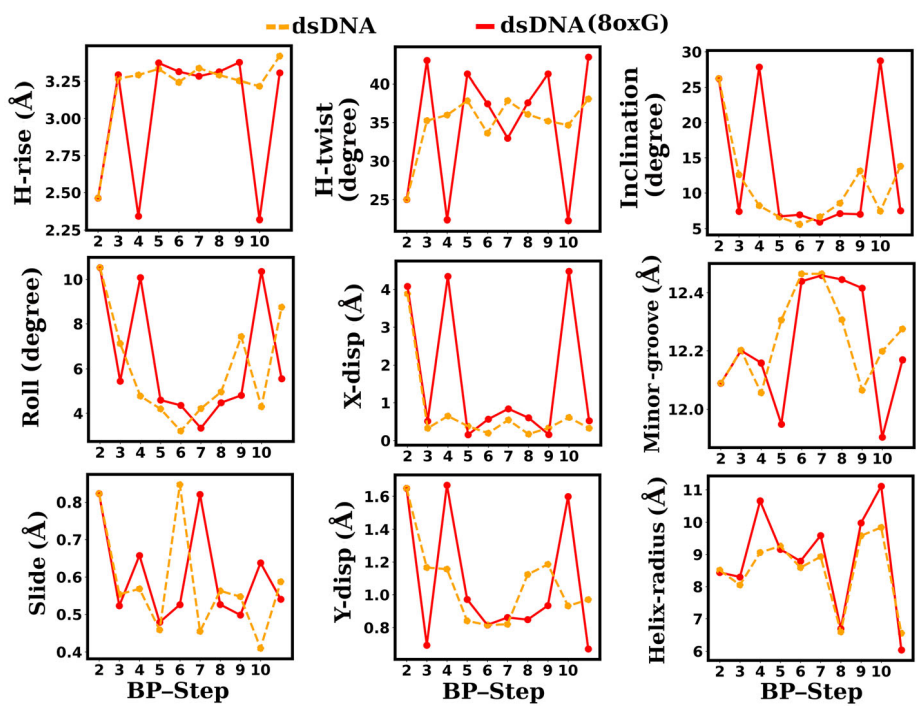

**Fig. 5** Average of absolute values of helical and groove parameters with respect to base pair step (BP-step), excluding end base pairs. For the clarity of plots, only native dsDNA and dsDNA(8oxG) parameters are shown, taking dsDNA(8oxG) as representative of the oxidized dsDNAs. These plots show that the changes in helical and groove parameters of oxidized dsDNA are concentrated at the local sites of oxidation

to fall into the normal C2'-endo range, discussed in sect. 4. A-form geometry of dsDNA or dsRNA have C3'-endo sugar puckers [71]. Trieb et al. [72] suggests the conformational changes of dsDNA within B-form geometry as a transition state to canonical A-form geometry or at least structurally very similar. So, we can presume that the oxidative damage of dsDNA may have caused transitions within the B-form which is discussed extensively in sect. 4

The H-rise and H-twist values plotted in Fig. 4 show that oxidized CG base pairs (8oxoG-C pairs) have a higher range of fluctuations than non-oxidized ones and indicate that the B-form dsDNA with oxidation has higher stretch and torsional flexibilities. The roll parameter has a contributory role in the duplexes' bending process. The more the variation in a duplex's roll parameter, the higher its bending flexibility. The smaller values of standard deviation of the roll parameter (plotted on Fig. 4) and the smaller bending angle (presented on Table 2) indicate that the B-form dsDNA with oxidation has higher bending stiffness. The calculated helical parameters substantiate the estimated mechanical properties described in earlier sections.

Along with the helicoidal parameters, we have computed the groove parameters to describe how the groove widths and helical radius of the dsDNA are changed with its oxidative damage. The values of few groove parameters (major groove width and helix radius) are included in Table 4 and in the Fig. 4. With the oxidation of dsDNA, the helical radius is found to be increased slightly. The helical radius for native dsDNA is $9.62 \pm 0.15$ Å and that of dsDNA(8oxG) (all guanines oxidized) is $9.83 \pm 0.20$ Å. The helical radius taken here is half of the distance between the phosphate atoms of two pairing bases. The consequence of the increase in helix radius is seen in the groove width also. The width of the major groove increases moderately from $19.15 \pm 0.78$ Å to $19.68 \pm 0.58$ Å when all the guanine bases are oxidized. Locally, the width of minor groove is found to decrease at the oxidation sites, which is depicted by Fig. 5.

The analysis of different helical and groove parameters showed that the oxidation causes effects on helical and groove properties as well, shifting the properties more or less from the native B-form characteristics, which is in agreement with the results of Peters et al. [20]. During base analog substitution, they found minor changes in global properties like persistence length, helical repeat and torsional rigidity but the data obtained from circular dichroism spectroscopy in their work showed some remarkable alters in helical geometry.

The helical parameters like slide, x-displacement, y-displacement and stretch are related to the dilation (lateral stretch) feature of dsDNA and their range (depicted by the standard deviation) should be related to their flexibility. The average values of slide, x-displacement, y-displacement and stretch are $0.18 \pm 0.71$ Å, $-0.42 \pm 1.71$ Å, $-0.06 \pm 1.34$ Å and $-0.065 \pm 0.122$ Å, respectively, for native dsDNA and $-0.46 \pm 1.27$ Å, $-1.79 \pm 3.03$ Å, $-0.13 \pm 1.94$ Å and $-0.067 \pm 0.143$ Å, respectively, for dsDNA(8oxG). The averages of absolute values of slide, x-displacement, y-displacement and stretch are $0.50$ Å, $1.27$ Å, $1.03$ Å and 0.10 Å , respectively, for native dsDNA and 0.63 Å, 2.53 Å, 1.25 Å and 0.13 Å , respectively, for dsDNA(8oxG). From these data, we can say that with oxidation of its bases, the dsDNA gets more flexible laterally and has high flexibility to dilate. The flexibility to change its lateral dimension certainly makes the structure flexible to extend and contract, lowering its stretch rigidity. The lateral extension of dsDNA is due to the repulsion between the negative phosphate and O8 atom of 8oxoG and the steric hindrance of O4$'$ of ribose and O8 of 8oxog [73], schematically shown in Fig. 7(b).

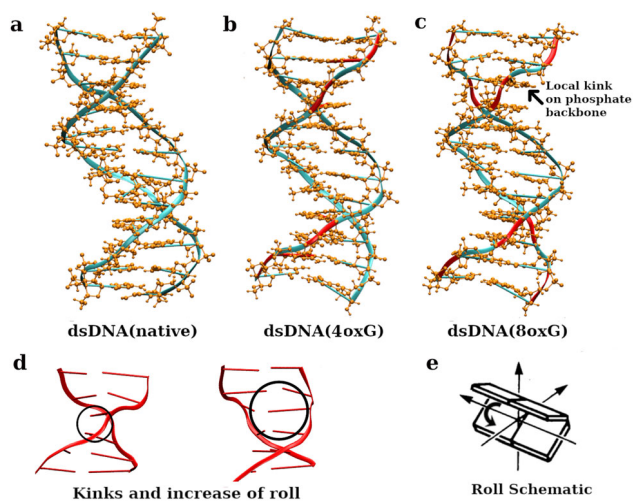

**Fig. 6** (**a–c**) Instantaneous snapshots of the duplexes at the end of 500 ns long MD simulations, which show local distortions (kinks) at the phosphate backbones of oxidation sites. The backbones are found to be distorted locally at the oxidation sites of the duplexes. (**d**) Local kinks and their effect on roll parameter. (**e**) Schematic of the roll parameter

Along with the helicoidal parameters averaged over all guanine-involved base pairs, we have computed the helicoidal parameters corresponding to each base pair step (BP-step). Figure 5 consists of the plots of the helical and groove parameters of native dsDNA and dsDNA(8oxG), for which we found significant changes with oxidation of the guanine bases. From these plots, we found the decrements on the following parameters: H-twist, H-rise and minor groove, at the oxidation sites. While computing the minor groove width by taking the average of all guanine-involved base-pair steps, it is found to be almost identical for oxidized and native dsDNA duplexes. However, when we are concerned with the values of individual base-pair steps, the minor groove is found to decrease at the oxidation sites, as shown in Fig. 5. It signifies that the minor groove is narrowed at the oxidation sites for oxidized duplexes, even though it is a global parameter. It may be due to the local distortions of the backbones, as shown in Fig. 6. Similarly, we found the increase for the following parameters: inclination, roll, stretch, slide and y-displacement, as well as helix radius, at those oxidation sites. From these plots, we can infer that the changes on helical and groove parameters are concentrated at the local sites of oxidation rather than on the whole length of the duplex.

The changes on the groove dimensions and the increase in inclination parameter can sometimes result local kinks on the backbones [23, 74, 75]. The increase of roll parameter at the local sites of oxidation is associated with such local kinks that occur at the backbones of the oxidized bases. The upper panel of Fig. 6 shows the local kinks occurred at the backbones of the oxidized duplexes and the lower panel of it describes the increase in roll parameter due to such local kinks. Such distortions on backbone of the dsDNA duplexes during the simulation can be seen in RMSF plots which is visible for the oxidized duplexes in Fig. 3(b). The bending of dsDNA duplex should be enhanced by the increase of roll parameter. However, the global bending is relatively smaller for oxidized duplexes than the native one. But locally, the roll parameter is found

**Fig. 7** (**a**) Phosphate linkages at the oxidation sites. The phosphate linkages show transitions from BI conformation to BII conformation at the oxidation sites of oxidized dsDNAs. (**b**) Water bridge between the phosphate oxygen O2P and O8 atom of 8oxoG, which is responsible for stabilizing the oxidized dsDNAs

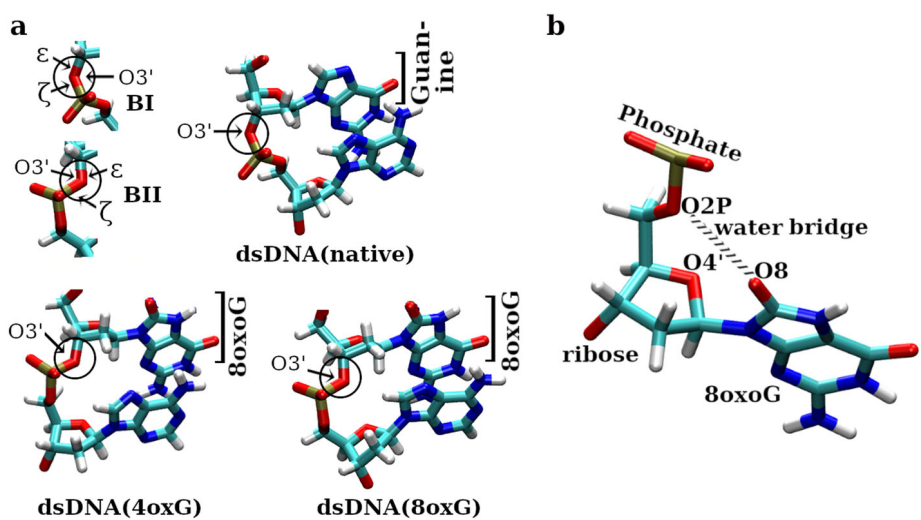

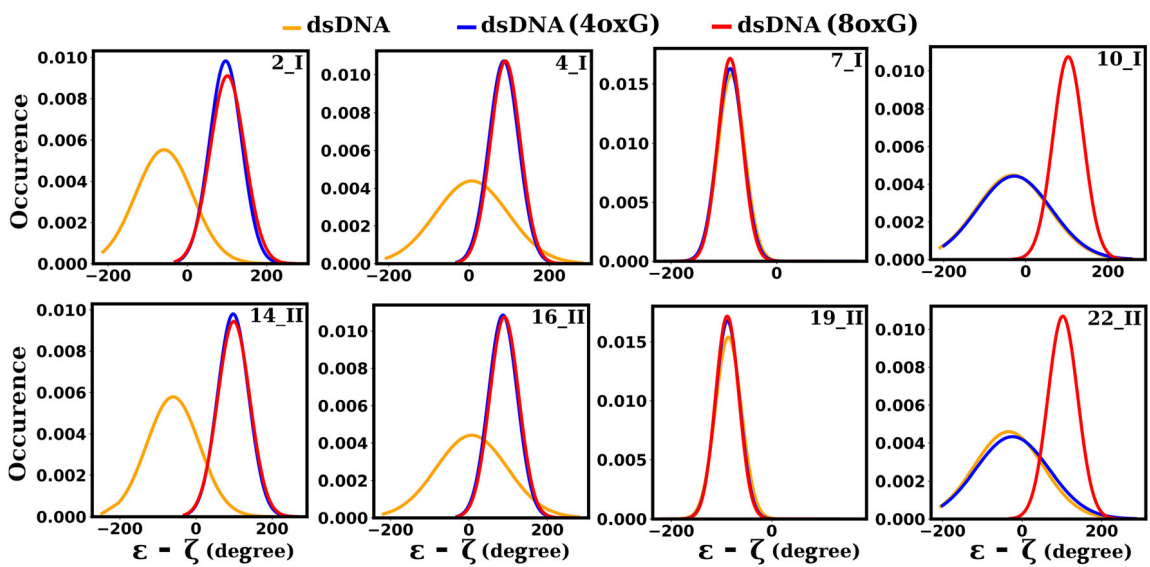

**Fig. 8** $\varepsilon - \zeta$ values for different nucleobases of the dsDNA duplexes. In the labeling at the top-right corner of each plot, first value represents residue ID of the nucleobase and second roman number represents the strand I or II. The $(\epsilon - \zeta \sim -90^o)$ condition indicates BI conformation and the $(\epsilon - \zeta \sim +90^o)$ condition indicates BII conformation of the backbone

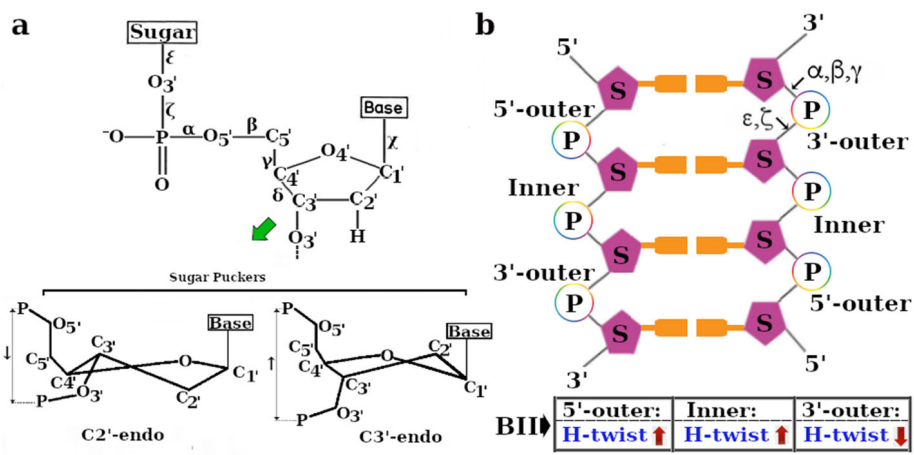

**Fig. 9** (**a**) Schematic for the backbone torsion angles. The black up/down arrow symbolizes increase/decrease of the phosphate to phosphate distance on the backbone. **b** Schematic for the changes in helical twist (H-twist) with respect to the position of BII substates. The red up/down arrow with H-twist symbolizes increase/decrease of the helical twist. Figures (**a**) and (**b**) are adapted from ref. [59, 84] and ref. [85], respectively

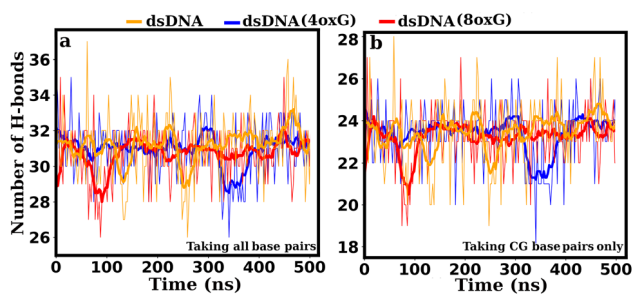

**Fig. 10** H-bonds between the base pairs of the two strands of dsDNA, dsDNA(4oxG) and dsDNA(8oxG). (**a**) Number of H-bonds taking into account all the base pairs. (**b**) Number of H-bonds considering CG base pairs only to be increased at the sites of oxidation (see Figs. 4 and 5 and Table 3). The increase of roll at the local sites of oxidation are, therefore, due to the local kinks on the phosphate backbones produced with the oxidative modification of the nucleobases.

## 4. Torsion angles analysis

The local changes in the helical and groove parameters at the oxidation sites of oxidized dsDNA prompted us to analyze the phosphate linkage at those oxidation sites. Figure 7(a) shows the phosphate linkage nearer the guanine base at the BP-step 4, one of the oxidation sites, where we can see the transitions of the phosphate linkage conformations from BI to BII in oxidized dsDNAs. We observed similar transitions at other oxidation sites (oxidized guanine bases) of the oxidized duplexes. BI and BII backbone

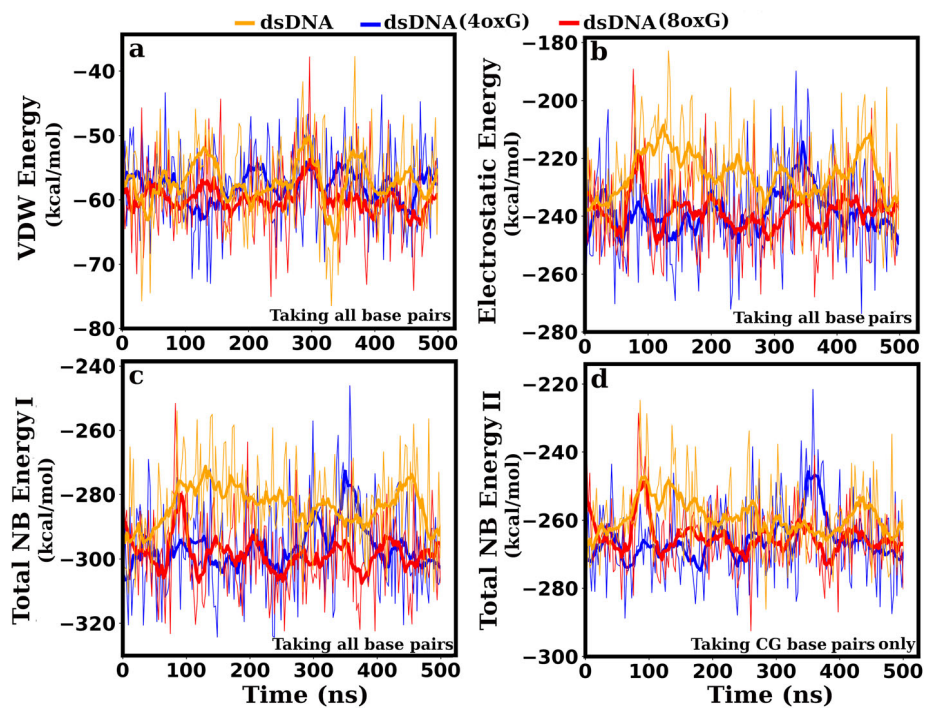

**Fig. 11** Non-bonded interaction energies between two strands of dsDNA, dsDNA(4oxG) and dsDNA(8oxG). (**a**) Van der Waal's (vdW) energy taking all base pairs. (**b**) Electrostatic energy taking all base pairs. (**c**) Total non-boned (NB) interaction energy taking all base pairs. (**d**) Total non-boned (NB) interaction energy taking CG base pairs only

**Table 4** Averages of some groove and helicoidal parameters

| System | Major groove width (Å) | Helix-radius (Å) | X-displacement (Å) | Slide (Å) |
| --- | --- | --- | --- | --- |
| dsDNA | 19.15 ± 0.78 | 9.62 ± 0.15 | −0.42 ± 1.71 | 0.18 ± 0.71 |
| dsDNA(4oxG) | 19.41 ± 0.84 | 9.69 ± 0.23 | −0.72 ± 2.12 | 0.16 ± 0.69 |
| dsDNA(8oxG) | 19.68 ± 0.58 | 9.83 ± 0.20 | −1.79 ± 3.03 | −0.46 ± 1.27 |

*The averages are taken from the values of the respective parameters of guanine-involved base pairs

conformations differ in the torsion angles $\epsilon$ and $\zeta$ which are, respectively, trans/g- in BI, i.e., $(\epsilon - \zeta = -ve)$ and g-/trans in BII, i.e., $(\epsilon - \zeta = +ve)$ [76–78], described more extensively in next paragraph. During our analysis, the twist-stretch coupling is found to be positive for all native as well as oxidized dsDNAs. As A-form geometries of dsDNA or dsRNA have opposite signs of twist-stretch couplings to their B-form geometries, we can say that the transitions seen in this work are not to A-form. From the phosphate linkage shown in Fig. 7(a), we can infer that the transitions seen in oxidized dsDNAs are from BI conformation to BII conformation, concentrated locally at the oxidation sites. To verify it, we have done the torsion angles analysis for the backbones at the local sites of oxidation.

The torsion angles for dsDNA duplexes are defined schematically in Fig. 9(a). The values of different backbone torsion angles for the duplexes are presented in Table St.2 of SI. In Table St.2 and Fig. 8, along with the oxidized bases, we have included some non-oxidized bases also to show the distinction between backbone torsion angle values of the oxidized and non-oxidized bases. For dsDNA(4oxG), residue IDs 2, 4, 14 and 16 are of the oxidized bases. For dsDNA(8oxG), residue IDs 2, 4, 10, 14, 16 and 22 are of the oxidized bases (excluding the end bases with residue IDs 12 and 24).

The backbone torsion angles are classified as gauche+ or g+ $(60 \pm 40^o)$, trans $(180 \pm 40^0)$ and gauche- or g- $(300 \pm 40^o)$ [79]. The sugar conformations are classified on the basis of phase angle $\phi$ as 'north' $(300-60^o)$, 'East' $(60-120^o)$, 'South' $(120-220^o)$ and 'West' $(220-300^o)$ [79, 80]. The North sugar conformation is characterized as C2'-exo if $\phi$ is in the range $300-360^o$ and C3'-endo if $\phi$ is in the range $360-60^o$. The East sugar conformation is characterized as C4'-exo if $\phi$ is in the range $60-90^o$ and C1'-exo if $\phi$ is in the range $90-120^o$. The South sugar conformation is characterized as C2'-endo if $\phi$ is in the range $120-180^o$ and C3'-exo if $\phi$ is in the range $180-220^o$. The West sugar conformation is characterized as C4′-endo if $\phi$ is in the range $220-270^o$ and C1′-endo if $\phi$ is in the range $270-300^o$. The C2′-endo of south sugar conformation is dominant for B-form of dsDNA and the C3′-endo of North sugar conformation is dominant for A-form of dsDNA [80, 81].

We found the sugar conformations in C2′-endo of South conformation and the $\alpha/\beta/\gamma$ torsion angles of the type g-/trans/g+ in both oxidized and non-oxidized bases of the duplexes (see Table St.2). It signifies that no duplexes have transition to A-form geometry but can have transition within different conformations of B-form geometry. The BI and BII conformations are characterized by the values of $\epsilon - \zeta$ such that BI conformation has values of $\epsilon - \zeta \sim -90^o$ with $\epsilon$ in trans and $\zeta$ in g- conformations and BII conformation has values of $\epsilon - \zeta \sim +90^o$ with $\epsilon$ in g- and $\zeta$ in trans conformations [76–79]. During BI to BII conformational transformation, the $\chi$ and $v_m$ values are increased [82]. These features can be seen clearly in the torsion angle values of oxidized bases presented in Table St.2 and Fig. 8. From these, we can say that the oxidized bases have conformational transitions from BI to BII within B-form geometry. Table St.2 shows the $\epsilon - \zeta$, $\chi$ and $v_m$ values of the oxidized bases as that of BII conformation, whereas that of non-oxidized bases (included only for illustration) as that of BI conformation. So, if all the bases of the duplexes are taken into consideration, the transition of the oxidized duplexes is of the hybrid type. For such hybrid conformation (BI and BII mixed), the values of helical parameters like slide, roll and twist are not significantly changed, unlike that of extreme BII conformation [83], which is the scenario observed in the results of our work as well.

Madhumalar and Bansal [82] classified BII conformations as weak, strong and hybrid types. In strong BII conformations, they report large twists, high positive slides and negative roll values. In weak BII conformations, these roll, twist and slide values are reported to be similar to the BI.BII hybrid type. They found BI.BII hybrid steps are more favorable than others. However, from the works of Hartmann et al. [86], Heddi et al. [87], Drsata et al. [88] and Zgarbova et al. [85], the changes in the helical twist are found to depend on the presence of BII substates. How the helical twist changes in the presence of BII substates is depicted in Fig. 9(b). The helical twist is found to increase when a BII substate is present on an inner phosphate [86, 87]. In contrast, the helical twist is found to reduce strongly when the BII substate is present on a 3′-outer phosphate [88]. Zgarbova et al. [85] found that the presence of BII at 5′-outer phosphate increases the helical twist but insignificantly. Zgarbova et al. [85], showed that the effect of 3′-outer phosphate is so significant that the overall helical twist is decreased when BII substates are present at different positions on a B-form dsDNA. In our case, the helical twist (taking guanine involved base pairs only) is found to decrease with oxidation of dsDNA and the backbone transitions from BI to BII conformations are local at the oxidation sites. Because the transitions are local at the oxidation sites, globally it can be taken as a hybrid type with positive $\epsilon - \zeta$ values at those oxidative damage sites. Similar transitions are reported in an MD simulation work by Ishida [73] and in the NMR experiments of Hoppins et al. [89] on single guanine oxidized dodecamer duplexes of DNA.

### 4.1. H-bonds and non-bonded energy analysis

The hydrogen bonds (H-bonds) between the base pairs of dsDNA help to assess the degree of dsDNA denaturation and its structural changes at the atomic level due to the oxidative damage. To check whether H-bonds between the base pairs are affected or not with oxidation of dsDNA, we computed the number of H-bonds between the Watson-Crick base pairs using the Hydrogen Bond plugin provided with the VMD [60], taking into account of all the base pairs as well as only CG base pairs. During H-bond analysis, the distance cut-off was taken 3.5 Å and the angle cut-off was taken $120^o$ as suggested by IUPAC [90]. From the H-bond analysis, we found no effect of oxidation on the hydrogen bonding between the base pairs. Taking all base pairs (see Fig. 10(a)), the H-bonds are found to fluctuate with an average value of 31 in each of the three dsDNA molecules. Similarly, taking CG base pairs only (see Fig. 10(b)), the H-bonds are found to fluctuate with an average value of 23 on all of them. In Dickerson-Drew dodecamer (d[CGCGAATTCGCG]) of double-stranded DNA, the H-bonds between the base pairs must be 32 in its normal condition and taking CG base pairs only there should be 24 such H-bonds. So, the calculated number of H-bonds show no significant reduction in their values from the expected ones. From this, it is apparent that oxidation only changes the helical and mechanical properties and does not deform the Watson-Crick base pair complementarity.

Through non-bonded (NB) energy analysis, taking all 12 base pairs (see Fig. 11(a-c)), the electrostatic energy is found to be -223.25 ± 15.40 kcal/mol for dsDNA, -235.29 ± 17.01 kcal/mol for dsDNA(4oxG) and -236.54 ± 15.67 kcal/mol for dsDNA(8oxG). The Van der Waals (vdW) energy is found to be -56.54 ± 4.71 kcal/mol for dsDNA, −57.17 ± 4.16 kcal/mol for dsDNA(4oxG) and -59.07 ± 4.36 kcal/mol for dsDNA(8oxG). And hence, the total NB energy (vdW and electrostatic) is − 279.79 ± 13.50 kcal/mol for dsDNA, -292.46 ± 18.98 kcal/mol for dsDNA(4oxG) and -295.62 ± 14.27 kcal/mol for dsDNA(8oxG). Along with it, we have computed the total NB energy taking CG base pairs only (see Fig. 11(d)) and found a similar trend that the total NB energy is -256.16 ± 16.83 kcal/mol for dsDNA, − 262.98 ± 17.35 kcal/mol for dsDNA(4oxG) and − 261.65 ± 18.41 kcal/mol for dsDNA(8oxG). The non-bonded energy analysis shows that dsDNA oxidation increases the non-bonded interaction between the bases. Taking CG base pairs only, although the total NB energy is found greater for the oxidized dsDNAs than the native dsDNA, the dsDNA(8oxG) has total NB energy insignificantly smaller than that of dsDNA(4oxG). The reason behind it may be due to the repulsion between the O8 oxygen atoms of the oxidized guanine (8oxoG) bases that persists when oxidized bases come closer in the DNA rungs during their symmetric presence as in dsDNA(8oxG). On oxidized nucleobases, the electron distribution and dipole moment are quite different from those of non-oxidized ones [91, 92]. The induced dipoles among canonical nucleobases with the presence of the O8 oxygen atoms of 8oxoG bases in oxidized dsDNAs are responsible for the increase in the electrostatic energy and pi-stacking of the nucleobases [93]. Another reason for the higher stability seen on oxidized dsDNA is the formation of a water bridge between phosphate oxygen O2P and O8 atom of 8oxoG. As the oxidized dsDNAs are found to be relatively stretched laterally, it provides space to form the water bridge between the phosphate and O8 of 8oxoG [73], schematically shown in Fig. 7(b). The attractive electrostatic interaction is found more significant than vdW energy with oxidation of the dsDNAs. The increase in the negative value of electrostatic interaction must be due to the presence of reactive oxygens on the oxidated guanine bases that makes strong interaction with the neighboring atoms. Thus from the results of NB energy analysis, we found no significant alterations on Watson-Crick geometries during the oxidation of nucleotide bases of dsDNA. Instead, the presence of reactive oxygens on guanines has improved the stability by increasing the attractive (-ve) electrostatic energy. Our results agree with the previously reported MD simulation results for 13-mer oxidized DNA by Cheng et al. [26]. In their results, oligomers containing G:C and 8oxoG:C pairs are found to adopt the Watson-Crick geometries in stable B-form duplexes, along with the change of the dynamic and equilibrium behavior of duplexes[26]. These changes of the dynamic and equilibrium behavior of duplexes are minor on global properties and more on helical properties [20].

## 5. Conclusions

In this work, we have presented the mechanical and helical properties of native and oxidized dsDNAs and explained those properties by microscopic structural parameters and thermodynamic calculations.

The persistence length, which is the measure of bending stiffness of dsDNA, is found to increase with its oxidative modification. The persistence length is 48.87 ± 1.81 nm for non-oxidized dsDNA and with oxidation of guanine bases, the persistence length increased to 58.61 ± 2.06 nm and 61.31 ± 2.44 nm for dsDNA(4oxG) (four guanine bases oxidized) and dsDNA(8oxG) (eight guanine bases oxidized), respectively, indicating that the oxidized dsDNA is relatively stiffer to bend than the non-oxidized dsDNA. It means that with oxidation, the bending flexibility of dsDNA gets suppressed. The stretch modulus, the measure of stretching flexibility, is 1239.26 ± 80.47 pN for non-oxidized dsDNA. With oxidation of its guanine bases, the stretch modulus became 800.68 ± 43.76 pN and 659.91 ± 35.82 pN for dsDNA(4oxG) and dsDNA(8oxG), respectively. Similarly, the torsional rigidity, the measure of the twisting stiffness, of non-oxidized dsDNA is 477.30 ± 11.16 pN.nm$^2$ and with oxidation of its guanine bases, the twisting rigidity became 424.51 ± 7.21 pN.nm$^2$ and 407.79 ± 13.32 pN.nm$^2$ for dsDNA(4oxG) and dsDNA(8oxG), respectively. It shows that the oxidized dsDNAs are more flexible to stretch and twist than the non-oxidized dsDNA. The higher flexibilities of oxidized dsDNAs to stretch and twist are related to their laterally stretched structures at the oxidized sites. The dsDNA duplex is laterally stretched due to oxidation and has high flexibility to dilate, which shows mechanical properties closure to A-form. However, as discussed above, the transition is not actually to the A-form but from BI to BII conformations, mainly localized at the sites of oxidative damage. The flexibility to stretch laterally makes the structure flexible to extend and contract, lowering its stretch rigidity.

The twist-stretch coupling of both non-oxidized and oxidized dsDNAs is found to be positive and signifies that the twist and stretch properties are positively correlated. That means increase/decrease of helical-twist should increase/decrease helical-rise of base pairs accordingly. The same (positive) sign of twist-stretch couplings for native dsDNA and oxidized dsDNAs also implies that the transitions of the oxidized duplexes are not to A-form geometry but within different conformations of B-form geometry.

Crookedness, the physical quantity related to the curvature, of dsDNA is found to be 0.18 ± 0.07 for non-oxidized dsDNA and became 0.43 ± 0.11 when all the guanine bases of it are oxidized. The greater value of crookedness is always associated with higher stretching flexibility. This is consistent with our finding that stretch modulus decreases with oxidation of the nucleotide bases of dsDNA.

The computation of helical parameters showed that helical-rise and helical-twist of dsDNA decrease with the oxidation of its bases. The inclination parameter is found to increase with the oxidation of the bases. The computation of groove parameters showed that the helix radius of dsDNA increases with the oxidation of its bases, whose direct consequence is on the increase of its major groove. In contrast, the minor groove width is observed to decrease at the oxidation sites. The changes in the groove dimensions and the increase in inclination parameter can sometimes result in local kinks on the backbones. The roll parameter increase at the local oxidation sites is associated with such local kinks that occur at the backbones of the oxidized bases. The effect of local kinks is also seen on the computation of RMSF and the RMSF curves of dsDNA(4oxG) and dsDNA(8oxG) are not as smooth as that of native dsDNA.

The non-bonded energy computation and hydrogen-bond analysis showed no significant denaturation of the bases of dsDNA due to their oxidation. Instead, the addition of oxygen atoms on the nucleotide bases increased the electrostatic energy between nucleobases, which results in the increase in the total non-bonded energy of the dsDNA due to oxidation.

Thus, the oxidative damage of dsDNA does not weaken the Watson-Crick geometry. Instead, strengthen it with the increase in electrostatic interaction. The oxidative damage changes the mechanical, helical and groove parameters of dsDNA and these changes are more significant on twisting and stretching flexibilities compared to bending stiffness. Similarly, compared to the mechanical or global properties, the changes are more significant on helical and groove properties.

## 6. Limitation of the study

It is worth mentioning here that 500 ns production runs using all-atom MD simulations may not be adequate to sample the global minimum energy configuration. Through analysis of the trajectories, we found that the oxidized dsDNA duplexes are better stabilized energetically than the native ones due to enhanced electrostatic interaction between the atoms of the nucleobases of oxidized duplexes. However, to make more quantitative statement about the stability, one needs to compute the relative free energy difference between the oxidized dsDNA and native dsDNA either performing enhanced sampling simulation or conducting relatively longer MD simulations to ensure better sampling.

Finally, we believe that the results and inferences of this work about the changes in mechanical properties and helical and groove parameters will help to advance the understanding of the alterations on biological as well as physical properties of DNA due to oxidative damage of its nucleotide bases.

**Supplementary Information** The online version contains supplementary material available at https://doi.org/10.1007/s12648-022-02299-y.

**Acknowledgements** We thank TUE-CMS, IISc, Bangalore, funded by DST, for providing the CPU hours and DAE, India, for financial support. SN acknowledges IISc for the institute RA fellowship.

**Declarations**

**Conflicts of interest** There are no conflicts to declare.